%% file: paper.tex
\documentclass[10pt, journal, compsoc]{IEEEtran}

\input{packages}
\input{macros}

\begin{document}

\title{Parallel and External-Memory Construction of\\Minimal Perfect Hash Functions with PTHash}

\author{Giulio Ermanno Pibiri --
Roberto Trani
\IEEEcompsocitemizethanks{
\IEEEcompsocthanksitem
Giulio Ermanno Pibiri is with
the Ca' Foscari University of Venice,
Dorsoduro 3246, 30123 Venice (Italy).
Roberto Trani is with the
HPC-Lab, ISTI-CNR,
Via G. Moruzzi 1, 56126 Pisa (Italy).\protect\\
E-mail: giulioermanno.pibiri@unive.it, roberto.trani@isti.cnr.it
\IEEEcompsocthanksitem
We thank Sebastiano Vigna for helpful comments on a draft of the article
and Martin Dietzfelbinger for claryifing some details about the analysis of the CHD
algorithm.
This work was partially supported by the European Union's Horizon Europe research
and innovation programme (EFRA project, Grant Agreement Number 101093026).
}
}

\input{abstract}

\maketitle

\IEEEdisplaynontitleabstractindextext

\IEEEpeerreviewmaketitle

\input{introduction}

\input{related_work}

\input{main}

\input{impl}

\input{experiments}
\input{conclusions}

\appendix

\newtheorem{lemma}{Lemma}
\input{proof_expected_runtime}

\vspace{-0.5cm}
\bibliographystyle{IEEEtran}
\bibliography{IEEEabrv,biblio}

\vspace{-1.4cm}
\begin{IEEEbiography}[{\vspace{-6mm}\includegraphics[height=1in,clip,keepaspectratio]{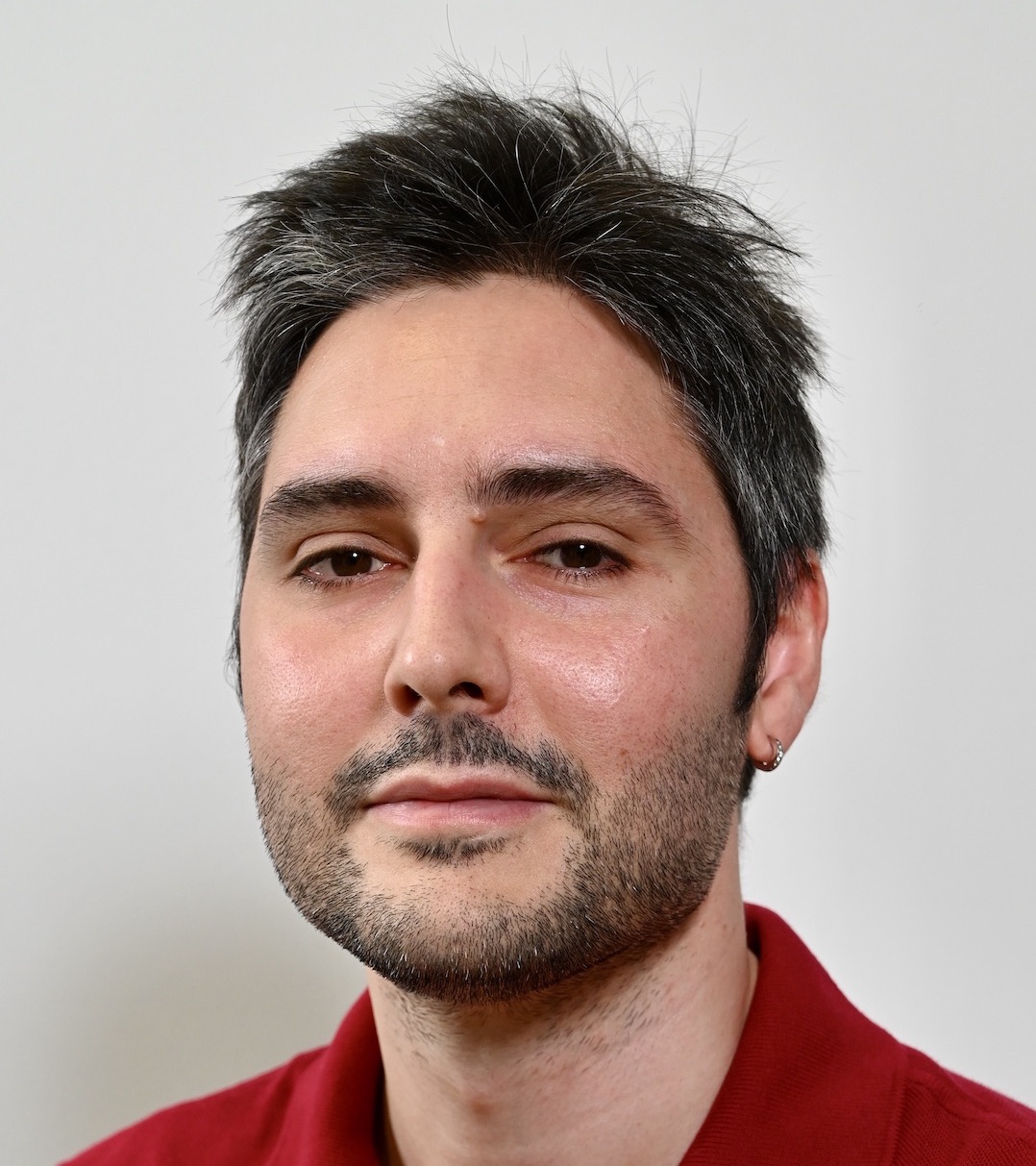}}]{Giulio Ermanno Pibiri}
is assistant professor at Ca' Foscari University of Venice, Italy.
He received a Ph.D. in Computer Science from the University of Pisa in 2019.
His research interests involve large-scale indexing, compressed data structures,
and algorithms, with applications to Information Retrieval and Computational Biology.
\end{IEEEbiography}

\vspace{-2.1cm}
\begin{IEEEbiography}[{\vspace{-6mm}\includegraphics[height=1in,clip,keepaspectratio]{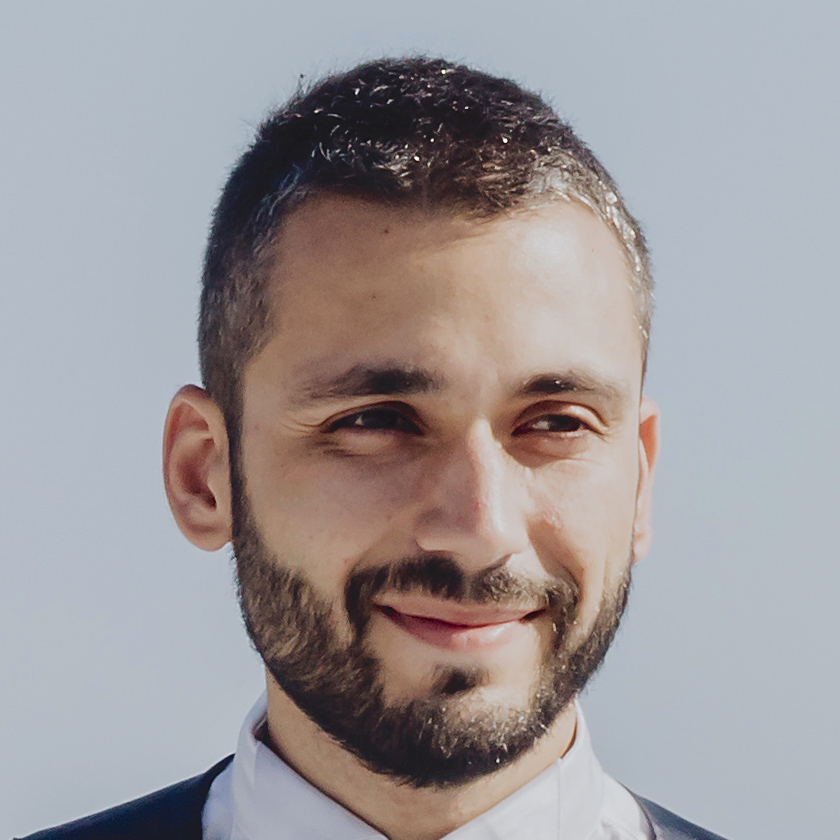}}]{Roberto Trani} received a Ph.D. in Computer Science from the University of Pisa, in 2020.
He is a postdoctoral researcher at the National Research Council of Italy (ISTI-CNR), within the High Performance Computing Laboratory, and his research interests are machine learning, algorithms, information retrieval, and high performance computing.
\end{IEEEbiography}

\end{document}

%% file: packages.tex
\usepackage{ragged2e} 
\usepackage{graphicx}
\usepackage{balance}
\usepackage{colortbl}
\usepackage{xcolor}
\usepackage{bm}
\usepackage{amsfonts}
\usepackage{url}
\usepackage{amsthm,amssymb,amsmath}
\usepackage{multirow}
\usepackage{booktabs}
\usepackage{makecell}
\usepackage[caption=false]{subfig}
\usepackage{wrapfig}
\usepackage{xcolor, soul}
\usepackage{siunitx}
\usepackage{enumitem}
\usepackage{rotating}
\usepackage{tablefootnote}
\usepackage[noend]{algpseudocode}
\usepackage[linesnumbered
            ,vlined]{algorithm2e}

\let\oldnl\nl
\newcommand{\nonl}{\renewcommand{\nl}{\let\nl\oldnl}}

\SetKwIF{If}{ElseIf}{Else}{{{if}}}{}{{else if}}{{{else}}}{}
\SetKwFor{For}{{for}}{}{}
\SetKwFor{While}{{while}}{}{}

\makeatletter
\newcommand{\removelatexerror}{\let\@latex@error\@gobble}
\makeatother

\usepackage{listings}

\usepackage[sc]{mathpazo}

%% file: macros.tex
\newtheorem{thm}{Theorem}

\newtheorem{fact}{Fact}

\DeclareMathOperator{\ExpectedValue}{\mathbb{E}}
\DeclareMathOperator{\Probability}{\mathbb{P}}
\DeclareMathOperator{\Binomial}{\mathtt{Binomial}}
\DeclareMathOperator{\Poisson}{\mathtt{Poisson}}

\newcommand{\mytablescale}{1.0} 
\newcommand{\mycaption}[1]{\caption{#1}}

\newcommand{\mymod}[1]{\,\textup{mod}{\,#1}}
\newcommand{\code}[1]{\mbox{\textbf{#1}}}
\newcommand{\var}[1]{\mbox{\emph{#1}}}
\newcommand{\bit}[1]{\mbox{\texttt{#1}}}
\newcommand{\func}[1]{\mbox{#1}}

\usepackage{xcolor, soul}
\usepackage[normalem]{ulem}
\newcommand{\remove}[1]{\bgroup\markoverwith{\textcolor{red}{\rule[1pt]{1pt}{2pt}}}\ULon{#1}}

\newcommand{\secs}{{seconds}}

\newcommand{\nspk}{{ns/key}}
\newcommand{\bpk}{{bits/key}}

\usepackage{pbox}
\usepackage{makecell}
\usepackage{tikz}
\usetikzlibrary{calc}

\newcommand{\method}[1]{\textup{{#1}}}
\newcommand{\pth}{\method{PTHash}}

\definecolor{tomato}{HTML}{E74C3C}
\definecolor{gray}{HTML}{D5D8DC}
\definecolor{darkgray}{HTML}{707070}
\definecolor{green}{HTML}{D5F5E3}

%% file: abstract.tex

\IEEEtitleabstractindextext{

\begin{justify}
\begin{abstract}
A function $f : U \to \{0,\ldots,n-1\}$ is a \emph{minimal perfect hash function}
for a set  $S \subseteq U$ of size $n$, if $f$ bijectively maps $S$
into the first $n$ natural numbers.
These functions are important for many practical applications
in computing, such as search engines, computer networks,
and databases.
Several algorithms have been proposed
to build minimal perfect hash functions that:
scale well to large sets,
retain fast evaluation time,
and take very little space, e.g., 2 -- 3 bits/key.
{\pth} is one such algorithm, achieving very fast
evaluation in compressed space, typically many times
faster than other techniques.
In this work, we propose a new construction algorithm for {\pth}
enabling:
(1) \emph{multi-threading}, to either build functions
more quickly or more space-efficiently,
and (2) \emph{external-memory processing}, to scale
to inputs much larger than the available internal memory.
Only few other algorithms in the literature share these
features, despite of their practical impact.
We conduct an extensive experimental assessment
on large real-world string collections
and show that, with respect to other techniques, {\pth} is competitive
in construction time and space consumption,
but retains 2 -- 6$\times$ better lookup time.


\end{abstract}
\end{justify}


}

%% file: introduction.tex

\section{Introduction}\label{sec:introduction}

\IEEEPARstart{M}{inimal} perfect hashing (MPH)
is a well-studied and fundamental problem in Computer Science.
It asks to build a data structure mapping the $n$ keys of a set $S$
out of a universe $U$ into the numbers $[n]=\{0,\ldots,n-1\}$.
In other words, the
resulting data structure realizes a ``one-to-one''
correspondence between $S$ and the integers in $[n]$. 
Such a function $f : U \rightarrow \{0,\ldots,n-1\}$
is called a
\emph{minimal perfect hash function} (MPHF henceforth)
for $S$.

MPHFs are useful in all those practical situations where
space-efficient storage and fast retrieval
from static sets is a primary concern.
In fact, they are employed in
compressed full-text indexes~\cite{belazzougui2014alphabet},
computer networks~\cite{lu2006perfect},
databases~\cite{chang2005perfect},
prefix-search data structures~\cite{belazzougui2010fast},
language models~\cite{
PibiriV19,StrimelRTPW20},
indexes for DNA~\cite{pibiri2022ISMB,pibiri2023algomb,pibiri2023lphash},
Bloom filters and their variants~\cite{broder2004network,fan2014cuckoo,graf2020xor},
and many other applications.



The most interesting aspect of the MPH problem
is that it ignores the behavior of $f$ on keys
that do \emph{not} belong to $S$, i.e.,
$f(x)$ can be any value in $[n]$ \emph{if} $x \in U \setminus S$.
Therefore, the MPHF data structure does not need to store the keys.
As a result, pioneer work on the problem has proved
a space lower bound of
$n\log_2(e)$ bits for the size of any MPHF~\cite{fredman1984storing,mehlhorn1982program},
which is approximately just 1.442 {\bpk}.
While it is difficult to come close to this space usage,
several practical algorithms exist that
take little space, i.e.,
2 -- 3 {\bpk}, retain very fast lookup time,
and scale well to large sets, e.g., billions of keys.

Our initial investigation on the problem
was motivated by the observation that MPHFs are usually
built once and evaluated many times, thus
making lookup time the most critical
aspect for the MPH problem,
provided that both construction time
and space usage are reasonably low.
To this end, we proposed {\pth}~\cite{pth_sigir},
which is significantly faster
at lookup time than other techniques
\emph{while} taking a similar memory footprint.
In our previous work, however, we only proposed
a sequential and internal-memory construction.

In this work, we extend the treatment of {\pth}
and consider two important algorithmic aspects:
\emph{multi-threading} and
\emph{external-memory processing}.
While multi-threading can be used to either
quicken the construction or build more space-efficient functions,
temporary disk storage can be used to scale to inputs much
larger than the available internal memory.
Only few other algorithms in the literature support these two features,
despite of their practical impact.


A simple and elegant solution to harness both aspects
is to partition the input and build, in internal memory,
an independent MPHF on each partition~\cite{botelho2013practical}.
This approach also offers good scalability as the
independent partitions that fit into internal memory
may be processed in parallel by multiple threads.
Very importantly, this solution is
valid for \emph{any} construction algorithm.
However, this approach imposes an indirection at lookup time
to identify the partition, i.e., the proper MPHF to query,
and additional space to store the offset of each partition.
Indeed, partitioned MPHFs are usually $30-50\%$ slower
and $20\%$ larger than their non-partitioned counterpart~\cite{botelho2013practical,belazzougui2014cache}.

\subsection*{Our Contribution}
Since the construction can either use one or multiple threads
and can either run in internal or external memory,
we consider the four possible construction settings:
sequential internal-memory, parallel internal-memory,
sequential external-memory, parallel external-memory.

We present a new construction
algorithm for {\pth} that overcomes
the overhead of the folklore partitioning approach and,
yet, easily adapts to all of the above four settings.
We conduct an extensive experimental assessment
over real-world string collections,
ranging in size from tens of millions to several billions of strings.
We show that {\pth} is competitive at building MPHFs
with the best existing techniques that also support
parallel execution and external-memory processing.
However, and very importantly,
{\pth} retains the best lookup time being 2 -- 6 times faster than other techniques.
Our C++ MPHF library is publicly available at
\url{https://github.com/jermp/pthash}.

\subsection*{Organization}
The article is structured as follows.
Section~\ref{sec:related_work} reviews all the practical approaches
for MPH and indicates which techniques support multi-threading
and/or use external memory.
Section~\ref{sec:main} presents
a new general {\pth} construction algorithm,
with Section~\ref{sec:impl} describing how
its design seamlessly adapts to the different settings
we consider.
In particular, Section~\ref{sec:int_mem} and~\ref{sec:ext_mem}
detail how the general construction works
in internal and external memory, respectively,
with multi-threading support (Section~\ref{sec:parallel_search}).
Section~\ref{sec:experiments} presents experimental
results, in comparison with the
approaches reviewed in Section~\ref{sec:related_work}.
We conclude in Section~\ref{sec:conclusions}.

%% file: related_work.tex

\section{Related Work}\label{sec:related_work}

Minimum perfect hashing has a long development history.
Our focus is on practical approaches that have been implemented and shown
to perform well on large key sets.
Indeed, we point out that some theoretical constructions,
like that by Hagerup and Tholey~\cite{hagerup2001efficient},
achieve the space lower bound of $n\log_2(e)$ bits,
but only work in the asymptotic sense, i.e., for $n$
too large to be of any practical interest.

Up to date, four ``classes'' of different, practical,
approaches have been devised to solve the problem, which we
describe below in chronological order of proposal.
Table~\ref{tab:notation} summarizes the notation
used to describe all algorithms while Table~\ref{tab:construction_bounds}
shows the theoretical construction times and the implementation aspects
of each algorithm.

\subsection*{Hash and Displace}
The \emph{hash and displace} technique was originally introduced
by Fox, Chen, and Heath~\cite{fox1992faster} in a work that was named FCH after them.
That work also inspired the development of {\pth}~\cite{pth_sigir}.
The main idea of the algorithm is as follows.

Keys are first hashed and mapped into
$\lceil cn / \log_2(n) \rceil$ \emph{non-uniform} buckets, for a given parameter $c > \log_2(e)$;
the buckets are then sorted and processed by falling size:
for each bucket, a displacement value $d_i \in [n]$ is determined
so that all keys in the bucket can be placed without
collisions to positions $(h(x) + d_i) \mymod n$,
using a random hash function $h$.
Lastly, the sequence of displacements $d_i$ is stored in compact form
using $\lceil \log_2(n) \rceil$ bits per value.
While the theoretical analysis suggests that by decreasing the number of buckets
it is possible to lower the space usage,
at the cost of a larger construction time,
in practice it is unfeasible to go below 2.5 {\bpk}
for large values of $n$.

In the \emph{compressed hash and displace} (CHD) variant by Belazzougui et al.~\cite{belazzougui2009hash},
keys are first \emph{uniformly} distributed to $\lceil n / \lambda \rceil$ buckets,
for a given parameter $\lambda \geq 1$.
The buckets are then sorted and processed by falling size:
for each bucket,
a pair of displacements $\langle d_0,d_1 \rangle$ is determined
so that all keys in the bucket can be placed without
collisions to positions
$(h_1(x) + d_0 h_2(x) + d_1) \mymod N$
using a pair of random hash functions $(h_1, h_2)$,
with $N=(1+\varepsilon)\cdot n$ slots, for a given parameter $\varepsilon > 0$.
Instead of storing a pair $\langle d_0,d_1 \rangle$
for each bucket, CHD stores the index of such pair in the sequence
$$\langle 0,0 \rangle,\ldots,\langle 0,N-1 \rangle,
\ldots,
\langle N-1,0 \rangle,\ldots,\langle N-1,N-1 \rangle.$$
The sequence of indexes is stored in compressed form
using an entropy coding mechanism with $O(1)$ access time~\cite{fredriksson2007simple}.

\begin{table}[t]
\centering
\caption{General notation (upper) and {\pth}-specific notation (bottom).\vspace{-3mm}}
\scalebox{\mytablescale}{\input{tables/notation.tex}}
\label{tab:notation}
\end{table}

\begin{table}[t]
\centering
\caption{Theoretical construction time and implementation details.\vspace{-3mm}}
\scalebox{\mytablescale}{\input{tables/construction_bounds.tex}}
\label{tab:construction_bounds}
\end{table}

\subsection*{Linear Systems}
In the late 90s, Majewski et al.~\cite{majewski1996family} introduced an algorithm
to build a MPHF exploiting a connection between
linear systems and hypergraphs.
Almost ten years later, Chazelle et al.~\cite{chazelle2004bloomier} proposed an analogous
construction in an independent manner.
The MPHF $f$ is found by generating a system of $n$ random equations in $t$
variables of the form
$$f(x) = w_{h_1(x)\mymod{t}} + \cdots + w_{h_r(x)\mymod{t}} \mymod{n}, \text{ for } x \in S,$$
where $\{ h_i \}_{i \in [r]}$ are $r$ random hash functions
and $\{ w_i \}_{i \in [t]}$ are $t$ variables whose values are in $[n]$.
Due to bounds on the acyclicity of random graphs,
the system can be triangulated with high probability and solved
in linear time by peeling the corresponding hypergraph
if the ratio between $t$ and $n$ is above a certain threshold $\gamma_{r}$.
The constant $\gamma_{r}$ depends on the degree $r$ of the graph
and attains its minimum for
$r=3$, where the value is $\gamma_{3} \approx 1.23$,
i.e., the system consists of $n$ equations of $3$ terms in $\lceil 1.23 n \rceil$ variables.
Recently, Dietzfelbinger and Walzer~\cite{DietzfelbingerW19} described a new family of
random hypergraphs, called \emph{fuse} graphs, that are peelable with high probability
even when the edge density is close to 1. Practically, this reduces the space
overhead of any solution based on linear systems from 23\% to 12\%.
Additional savings are possible using $r>3$ hash functions at the expense of
increasing the construction time.

Belazzougui et al.~\cite{belazzougui2014cache} proposed a cache-oblivious implementation
of the previous algorithm suitable for external memory construction
and named EMPHF.
The algorithm employs a compact representation
of the incidence lists of the hypergraph, which is based
on the observation that it is not necessary to store actual edges.
Instead, all vertices in the same position can be XORed together,
hence a constant amount of memory per node is retained.

Genuzio et al.~\cite{genuzio2016fast,genuzio2020fast} (GOV) demonstrated practical
improvements to the Gaussian elimination technique,
which is used to solve the linear system.
The improvements are based on broadword programming techniques.
The authors released a very efficient implementation
of the algorithm that scales well using external memory
and multi-threading.

\subsection*{Fingerprinting}
M\"uller et al.~\cite{muller2014retrieval} introduced
a technique based on fingerprinting.
The general idea is as follows.
All keys are first hashed in $[n]$ using a random hash function,
then collisions are recorded using a bitmap $B_0$ of size $n_0=n$.
In particular,
keys that do not collide have their position
in the bitmap marked with a \bit{1};
all positions involved in a collision are marked with
a \bit{0} instead. If $n_1 > 0$ collisions are produced,
then the
same process is repeated recursively for the $n_1$
colliding keys using a bitmap $B_1$ of size $n_1$.
All bitmaps, called ``levels'', are then concatenated together
in a bitmap $B$.
The lookup algorithm keeps hashing the key level by level
until a \bit{1} is hit, say in position $p$ of $B$.
A constant-time ranking data structure~\cite{jacobson1989space}
is used to count the number of \bit{1}s in $B[0..p]$
to ensure that the returned value is in $[n]$.
On average, only $1.56$ levels are accessed
in the most succinct setting~\cite{muller2014retrieval}, which takes 3 {\bpk}.


Limasset et al.~\cite{limasset2017fast} provided an implementation
of this approach, named BBHash, that is very fast in construction
and scales to very large key sets using multi-threading and external-memory processing.
Furthermore, no auxiliary data structures are needed during
construction except the levels themselves,
meaning that the space consumed in the process
is essentially that of the final MPHF.
The authors also introduced a parameter $\gamma \geq 1$ to speedup
construction and lookup time by using bitmaps that are
$\gamma$ times bigger at any level.
This clearly reduces collisions and, thus, the average
number of levels accessed per lookup.
However, the larger $\gamma$, the higher the space consumption.

\subsection*{Recursive Splitting}
Esposito et al.~\cite{esposito2020recsplit} proposed
a \emph{recursive splitting} algorithm (RecSplit)
to build very succinct functions, e.g., 1.80 {\bpk}, in expected linear time.
Also, it provides fast lookup time in expected constant time.
The authors first observed that, for very small sets,
it is possible to find a MPHF simply by brute-force searching
for a bijection with a suitable codomain.
Then, the same approach is applied in a divide-and-conquer manner
to solve the problem on larger inputs.

Given two parameters $b$ and $\ell$, the keys are divided into buckets
of average size $b$ using a random hash function.
Each bucket is recursively
split until a block of size $\ell$ is formed
and for which a MPHF can be found using brute-force, hence forming
a rooted tree of splittings.
The parameters $b$ and $\ell$ provide different space/time trade-offs.
While providing a very compact representation,
the evaluation performs one memory access for each level of the tree
that penalizes the lookup time.

The authors indicate that the approach is amenable
to good parallelization as the buckets may be processed
by multiple threads in parallel.
However, at the time of writing\footnote{When this paper was under revision, Bez et al.~\cite{bez2022high} proposed a GPU parallelization of the RecSplit algorithm.}, there is no public parallel implementation of this algorithm.

%% file: tables/notation.tex

\begin{tabular}{c l}

\toprule


$S$ & set of keys \\

$n$ & number of keys in $S$, $n = |S|$ \\

$[n]$ & $\{0,\ldots,n-1\}$ \\

$f$ & MPHF built from $S$ \\

$f(x)$ & a value in $[n]$, the result of computing $f$ on the key $x$ \\

\midrule

$h$ & a fully random hash function\tablefootnote{
We assume to have a family of \emph{fully random independent hash functions}
that can be evaluated in $O(1)$ time
from which we select $h$.
} \\

$c$ & value that trades search efficiency for space effectiveness \\

$\alpha$ & value in $(0,1)$, used to define the search space $\left[ N=\lceil n/\alpha \rceil \right]$ \\

$m$ & number of buckets used for the search, $m = \lceil cn / \log_2\left(n\right) \rceil$ \\

$L$ & largest bucket size \\

$P$ & pilots table \\

$x \oplus y$ & bitwise XOR between hash codes $x$ and $y$\\

\bottomrule
\end{tabular}

%% file: tables/construction_bounds.tex

\begin{tabular}{l c c c}

\toprule

\multirow{2}{*}{Algorithm}
&
\multirow{2}{*}{Construction Time}
&
\multicolumn{2}{c}{Implementation}
\\


&& multi-thr. & ext. mem. \\

\midrule

{\pth}      & $O\big(n^{1+\Theta(\alpha/c)}\big)$ & $\checkmark$ & $\checkmark$ \\
FCH         & not reported && \\
CHD         & $O\big(n\cdot(2^{\lambda}+(1/\varepsilon)^{\lambda})\big)$ && \\
EMPHF, GOV  & $O(n)$ &              & $\checkmark$ \\
BBHash      & $O(n \cdot e^{1/\gamma})$ & $\checkmark$ & $\checkmark$ \\
RecSplit    & $O(n)$ && \\

\bottomrule
\end{tabular}

%% file: main.tex

\section{A General Construction}\label{sec:main}

This section presents a general construction algorithm
for {\pth}, based on a few abstract steps, that is suitable
for both the parallel and external-memory settings.
While the focus of this section is on the algorithm
rather than on
the specific implementation, Section~\ref{sec:impl}
will detail how the proposed construction
can be easily implemented in practice to support the different settings.

At a high level, the construction first hashes and maps the keys into non-uniform buckets,
then processes the buckets in decreasing order of size:
for each bucket, an integer $k_i$ -- the \emph{pilot} of the $i$-th bucket --
is determined so that all keys in the bucket can be placed without collisions
to positions
$$f(x) = (h(x) \oplus h(k_i)) \mymod{n}$$
where $h$ is a random hash function\footnote{
To ensure independence between the hash codes, we would require \emph{two distinct} hash functions, one for the keys and one for the pilots.
Here, we re-use the same function for simplicity.
}.
Lastly, the sequence of such pilots, one for each bucket,
is materialized in a \emph{pilots table} (PT),
indicated with $P$ throughout the paper.
The table is stored in \emph{compressed} format
using a compressor supporting random access to $P[i]$
in $O(1)$ worst-case time.

\begin{figure}[t]
\removelatexerror
\SetArgSty{textnormal}
\begin{algorithm}[H]
\input{algs/framework}
\mycaption{The general construction algorithm for a set $S$ of $n$ keys,
and parameters $c$ and $\alpha$.
\label{alg:framework}}
\end{algorithm}
\end{figure}

Besides the input set $S$, the construction employs two
parameters, $c > \log_2(e)$ and $0 < \alpha < 1$,
respectively affecting the number of buckets
and the size of the search space --
refer to the bottom part of Table~\ref{tab:notation}
for the notation used to describe {\pth}.
The general construction algorithm is composed of
three steps, namely \Call{Map}{}, \Call{Merge}{}, and \Call{Search}{},
followed by an \Call{Encode}{} step that compresses the table as anticipated before.
These steps, explained in the following, are summarized in Alg.~\ref{alg:framework}.

\subsection{Map}\label{sec:map}

\begin{figure}[t]
\centering

\subfloat[]{
   \includegraphics[scale=0.57]{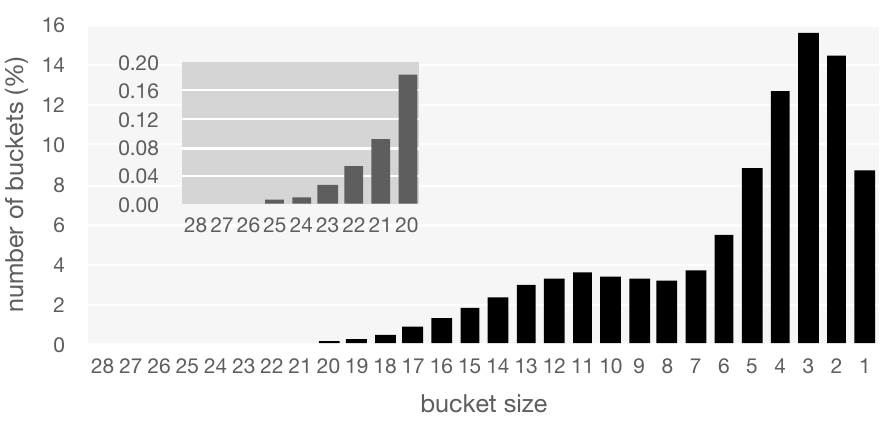}
   \label{fig:bucket_size_distribution}
}

\subfloat[]{
   \includegraphics[scale=0.57]{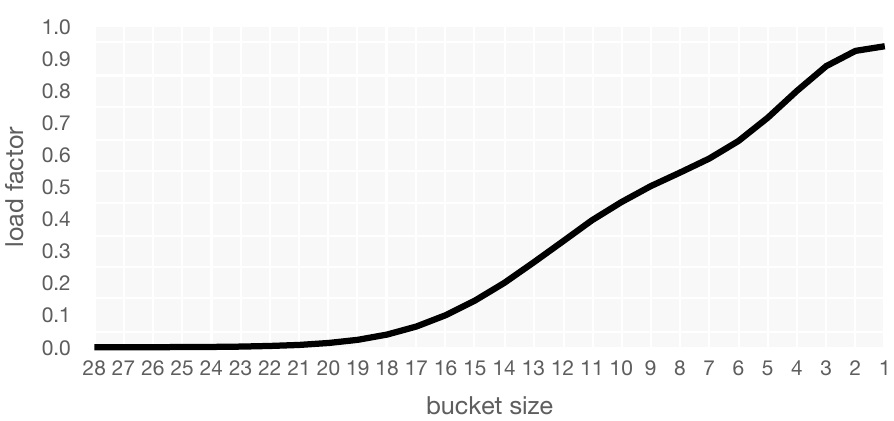}
   \label{fig:load_factor}
}

\caption{Bucket size distribution and
load factor as buckets are processed by decreasing size order,
for $n=10^6$ keys, $\alpha=0.94$, and $c=3.5$.
\label{fig:bucket_size_distribution_load_factor}}
\end{figure}

The first step aims at mapping each key of $S$
to a fixed number of bytes using a random
hash function $h$.
During this step, keys are ``logically'' associated to one of the
$m = \lceil cn / \log_2(n) \rceil$ buckets
$\{B_0,\ldots,B_{m-1}\}$,
that are indicated with the integers in $[0..m-1]$.
Specifically, the \emph{bucket identifier} for a key $x$ is
obtained via the function
$$
\func{bucket}(x) =
   \left\{
\begin{array}{ll}
      h^{\prime}(x)\mymod{p_2} & h^{\prime}(x)\mymod{n} < p_1 \\
      h^{\prime}(x) \mymod{(m-p_2)} + p_2 & \mbox{otherwise} \\
\end{array}
\right.
$$
where $h^{\prime}$ is a random hash function, like $h$, while
$p_1$ and $p_2$ are two thresholds set to,
respectively, $\frac{6}{10}n$ and $\frac{3}{10}m$ so that the mapping
of keys into buckets is \emph{uneven}:
roughly 60\% of the keys are mapped into 30\% of the
buckets.
This process induces a skewed distribution of the bucket sizes,
as depicted in the example of Fig.~\ref{fig:bucket_size_distribution}.
The example -- for $n=10^6$ keys 
and $c=3.5$ --
shows the percentage of buckets that have size equal to $|B|$,
for $|B|=L,\ldots,1$, with $L=28$ being the largest bucket size.
As evident, large buckets are very few while most of them are small.
Such skewed distribution of keys into buckets
is very important for the \emph{efficiency} of the
\Call{Search}{} step and for the final compression effectiveness,
as we are going to motivate in Section~\ref{sec:search}.

For each key $x \in S$ we generate an integer pair
$$
\langle \var{id} , \var{hash} \rangle =
\langle \func{bucket}(x), h(x) \rangle
$$
and divide the $n$ pairs into $K$ \var{blocks},
each of (roughly) equal size.
The pairs in each block are sorted by, first, the
\var{id} component, then by \var{hash}.
Blocks are formed either because: keys are evenly
distributed among $K$ threads and processed
in parallel in internal memory, or
a block of pairs is flushed to disk when internal
memory is exhausted. We will better explain these
two scenarios in Section~\ref{sec:int_mem}
and~\ref{sec:ext_mem}.

The \Call{Map}{} step takes $O(n \log(n/K))$ expected time
using quick-sort and
assuming the computation of $h(x)$
and $h^{\prime}(x)$ to take $O(1)$ time.
If we use $u$-bit hash codes, the temporary space
used during this step
is $n\cdot(\lceil\log_2(m) \rceil + u)$ bits.

\subsection{Merge}\label{sec:merge}

The $K$ blocks output by the previous step
are then merged to group together
all the hash codes of each bucket.
Specifically, the pairs are merged by their \var{id}
component first, then by \var{hash}. In this way we are also able to check
that all keys were correctly hashed to distinct hash codes%
\footnote{
Any MPHF construction fails in case of hash collisions.
However,
the failing probability is very low and depends only on $n$ and $u$. Indeed,
we never had collisions during
our experiments using $u=64$ bits for both $h(x)$ and $h^{\prime}(x)$.
}.
Moreover, since pairs having the same \var{id} are also
sorted by \var{hash}, checking for duplicate keys
is very efficient as it can be performed with a single scan.

During this step we also accomplish to
sort the buckets by non-increasing size
%
as follows.
We allocate $L$ buffers.
Suppose the merge for a bucket is complete
and the bucket contains $k$ hash codes.
We append its identifier and the $k$ hash codes
to the $k$-th buffer
so to have all the buckets
of the same size written contiguously.
Once the merge is complete,
the $K$ input blocks with the pairs are destroyed.

The output of this step is the set of such $L$
buffers, so that the next step, \Call{Search}{},
has to logically process the buffers from index
$L-1$ down to index $0$.
We indicate the $L$ buffers with the abstraction
\var{buckets} in Alg.~\ref{alg:framework}.
This abstract collection \var{buckets}
must only support \emph{sequential iteration}
in the wanted bucket order (from largest to smallest bucket size).
Sequential iteration is an access pattern
of crucial importance
as to avoid expensive memory reads
for the buckets, especially in the external-memory setting.
Each element of the collection is a \var{bucket} --
an object made by an array of \var{hashes},
plus a unique identifier $0 \leq \var{id} \leq m-1$.

The overall time complexity of \Call{Merge}{} is $O(n \log K)$
using a min-heap data structure of size $O(K)$ memory words
to merge the blocks.
Since each bucket in the $k$-th buffer is a contiguous span of $k+1$ integers,
for $0 \leq k < L$, the total space taken by the merge step is
$m \lceil \log_2(m) \rceil + nu + O(L)$ bits,
where the last term is negligible
as $L$ is very small compared to $n$,
e.g., $L < 40$ in our experiments with billions of keys.

\begin{figure}[t]
\removelatexerror
\SetArgSty{textnormal}
\begin{algorithm}[H]
\input{algs/seq_search}
\mycaption{The \textsc{Search} algorithm for an input collection of $m$ \var{buckets}
in a search space of size $N= \lceil n/\alpha \rceil$.
\label{alg:seq_search}}
\end{algorithm}
\end{figure}

\subsection{Search}\label{sec:search}

The \Call{Search}{} step is the core of the whole construction and
it is illustrated in Alg.~\ref{alg:seq_search}.
The algorithm keeps track of
occupied positions in the search space
$\left[ N=\lceil n/\alpha \rceil \right]=\{0,\ldots,N-1\}$
using a bitmap of $N$ bits, $\var{taken}[0..N-1]$,
that initially are all set to \bit{0}.
For each $\var{bucket} = \var{buckets}[i]$,
we search for an integer $k_i$, called pilot of the \var{bucket},
satisfying the condition $\var{taken}[p] = \bit{0}$
for all positions $p = (h(x) \oplus h(k_i)) \mymod{N}$
assigned to the hashes of the bucket keys $h(x) \in \var{bucket}.\var{hashes}$.
If the search for $k_i$ is successful, i.e.,
$k_i$ places all hashes of $\var{bucket}$
into different unused positions of \var{taken}
(checks of the lines 13 and 18),
then $k_i$ is saved in the pilots table $P$
(line 21)
and the positions are marked as occupied
via $\var{taken}[p]=\bit{1}$ (line 22);
otherwise, the next integer $k_i + 1$ is tried.

At this stage of the description,
$P$ is modeled as an abstract collection of pairs
$\langle \var{id}, \var{pilot} \rangle$.
The collection is then sorted by \var{id} to
materialize the final pilots table
(line 24). In this way, the code of \Call{Search}{}
can adapt to either work in internal or external memory.

The positions $p$ (line 12)
and pilots $k_i$ are 64-bit integers.
The \Call{Search}{} algorithm consumes
$N$ bits for the bitmap \var{taken},
plus $64 \cdot L$ bits for the array \var{positions}
because at most $L$ positions can be added to it,
plus $m\cdot(\lceil \log_2(m) \rceil + 64)$ bits
for \var{P}.

\subsection*{Expected Pilot Value and Search Time}
Each pilot $k_i$ is a random variable taking value $v \in \{0,1,2,\ldots\}$
with probability depending on the
current \emph{load factor} of the bitmap \var{taken}
(fraction of bits set).
It follows that $k_i$ is \emph{geometrically distributed}
with success probability $p_i$
being the probability of placing all keys
in $B_i$ without collisions.
Let $\alpha(i)$ be the load factor of the bitmap
after buckets $B_0,\ldots,B_{i-1}$ have been processed, that is
$
\alpha(i) = \frac{\alpha}{n}\sum_{j=0}^{i-1}|B_j|
$,
for $i = 1,\ldots,m$
and $\alpha(0)=0$ for convenience (empty bitmap).
Since the keys are displaced using a fully random hash function, the probability $p_i$ can be modeled as
$
p_i = (1 - \alpha(i))^{|B_i|}
$.


\begin{fact}
The expected number of trials per bucket is
\begin{equation}\label{fact:num_trials}
\ExpectedValue[k_i]+1 = 1/p_i = (1 - \alpha(i))^{-|B_i|}.
\end{equation}
\end{fact}
\begin{proof}
Since $k_i$ is geometrically distributed, $\Probability(k_i = v) = p_i(1 - p_i)^v$
which is the probability of succeeding after $v$ failures ($v+1$ trials).
Hence, $\ExpectedValue[k_i]=1/p_i - 1$.
\end{proof}

Fact~\ref{fact:num_trials} explains why the expected pilot value
-- thus, the expected number of trials per bucket --
is small on average.
In fact, \Call{Search}{} processes the buckets in decreasing size order
and the load factor is very small
when the exponent $|B_i|$ is large, i.e., during the initial phase of the process.
Then, as buckets are processed, the load factor grows
and the exponent $|B_i|$ shrinks.
Fig.~\ref{fig:load_factor} shows an example of how the load factor grows
due to all buckets of size $\leq |B|$, for $|B|=28,\ldots,1$.
Note the very slow growth rate of the load factor
-- e.g., it is smaller than 0.5 for $|B_i| \geq 10$
and higher than 0.9 only for $|B_i| \leq 2$.

Lastly, we give the following theorem
which relates the performance
of \Call{Search}{}, in terms of expected CPU time,
to the parameters $c$ and $\alpha$.
See the Appendix for a proof.

\begin{thm}\label{thm:search}
The expected time of \Call{Search}{} (Alg.~\ref{alg:seq_search}),
for $n$ keys and parameters $c > \log_2(e)$
and $0 < \alpha < 1$,
is $O(n^{1+\Theta(\alpha/c)}).$
\end{thm}

\subsection*{Minimal Output}
The search space is
$N=n/\alpha > n$ since $\alpha < 1$.
It makes the search for pilots faster by
lowering the probability of hitting already taken positions
in the bitmap.
However, we now need to build a mapping of the $n$ keys
from $[N]$ to $[n]$ to make $f$ minimal.

Suppose $F$ is the list of unused positions \emph{up to},
and including, position $n-1$.
Then, there are $|F|$ keys
that can fill such unused positions
which are mapped to positions $p_i \geq n$.
To this end, we materialize an array $\var{free}[0..N - n - 1]$,
where
$\var{free}[p_i - n] = F[i]$, for each $i=0,\ldots,|F|-1$.
It allows us to easily map the keys ending outside $[n]$
to distinct unused positions of $[n]$.
To compute the (uncompressed)
\var{free} array we need $\Theta(n)$ time
and $64\cdot(N - n)$ bits.

\subsection{Encode}\label{sec:encode}

The {\pth} data structure stores the two compressed tables,
$P$ and \var{free}.
For the \var{free} array, we always use
Elias-Fano~\cite{Fano71,Elias74},
noting that it only takes
$$
(N-n)(\lceil\log_2(n/(N-n))\rceil +2+o(1)) \text{ bits.}
$$
The pilots table $P$, instead,
can be compressed using \emph{any} compressor
for integer sequences that supports constant-time random access
to its $i$-th integer $P[i]$.
It is also desirable that compressing $P$ runs in linear time,
i.e., $\Theta(m)$ time.
We thus only consider compressors with this complexity
in the article.

Therefore, we have three degrees of freedom
for the tuning of {\pth},
namely the choice of
\begin{itemize}
\item the compressor for $P$,
\item the size of the search space, tuned with $\alpha$, and
\item the number of buckets, tuned with $c$,
\end{itemize}
that allow us to obtain different space/time trade-offs.
We point the interested reader to our previous paper~\cite{pth_sigir}
for an overview and discussion of the achievable trade-offs.
For example, a good balance between space effectiveness
and lookup efficiency is obtained using
the \emph{front-back dictionary-based} encoding
(D-D)~\cite{pth_sigir}, with $\alpha=0.94$
and $c=7.0$.
More compact representations can be obtained, instead,
by using Elias-Fano (EF) on the prefix-sums of $P$
at the price of a slowdown at query time.
We are going to use both configurations
in the experiments as reference points.

\begin{figure}[t]
\removelatexerror
\SetArgSty{textnormal}
\begin{algorithm}[H]
\input{algs/P_i}
\mycaption{Algorithm for retrieving the $i$-th element
of the pilots table when represented in partitioned compact (PC)
form. The notation $x = B[a,b]$, $a < b$, means that
$b-a$ bits are read from the bitmap $B$ starting at position $a$
and written to an integer value $x$.
\label{alg:P_i}}
\end{algorithm}
\end{figure}

\subsection*{Partitioned Compact Encoding}
Here, we propose another compressed
representation that supports constant-time random access,
and named \emph{partitioned compact} (PC) encoding hereafter.
The pilots table is divided into $\lceil m/b \rceil$ blocks of
size $b$ each, except possibly the last one (in our implementation
we use $b=256$).
For each block, we compute the minimum bit-width $w_i$
necessary to represent its maximum element,
i.e., $w_i = \lceil \log_2(\var{max}+1) \rceil$
(if $\var{max}=0$, then $w_i$ is set to 1),
so that
every integer in the block can be represented using $w_i$ bits.
The whole pilots table is represented by concatenating
the representations of all blocks in a bitmap $B$ that takes
$O(1) + b\sum_{i=0}^{\lceil m/b \rceil - 1} w_i$ bits.
To support random access, we also materialize a support array $W[0..\lceil m/b \rceil]$
containing the offset of all blocks,
i.e., $W[0]=0$ and $W[i] = \sum_{i=0}^{j-1} w_i$,
for $1 \leq i \leq \lceil m/b \rceil$.
The algorithm for retrieving $P[i]$ is illustrated in Alg.~\ref{alg:P_i}:
it requires two memory accesses (one to $W$, the other to $B$)
and no branches, thus it retains a fast evaluation.
The space effectiveness of the encoding is expected to be
good for the same reasons why front-back compression
works well~\cite[Section 4.2]{pth_sigir}:
the entropy of $P$ is lower at the front and higher at the back,
with a smooth transition between the two regions.
This means that the bit-widths $\{w_i\}$ tend to generally
increase when moving $i$ from $0$ to $\lceil m/b \rceil - 1$.
Indeed, the PC encoding can be considered as a generalization
of front-back compression.

%% file: algs/framework.tex
\DontPrintSemicolon
\SetAlgoCaptionSeparator{\footnotesize\normalfont\sffamily{.}}
\SetAlgorithmName{\footnotesize\normalfont\sffamily{Alg.}}{}
\SetAlCapFnt{\footnotesize\normalfont\sffamily}
\SetAlCapNameFnt{\footnotesize\normalfont\sffamily}
\SetKwBlock{Begin}{}{}
\Begin(\Call{Build}{$S$, $c$, $\alpha$})
{
    $\var{blocks} = \Call{Map}{S,c}$ \\
    $\var{buckets} = \Call{Merge}{\var{blocks}}$ \\
    $P = \Call{Search}{\var{buckets},n/\alpha}$ \\
    $\Call{Encode}{P}$ \\
}

%% file: algs/seq_search.tex
\DontPrintSemicolon
\SetAlgoCaptionSeparator{\footnotesize\normalfont\sffamily{.}}
\SetAlgorithmName{\footnotesize\normalfont\sffamily{Alg.}}{}
\SetAlCapFnt{\footnotesize\normalfont\sffamily}
\SetAlCapNameFnt{\footnotesize\normalfont\sffamily}
\SetKwBlock{Begin}{}{}
\Begin(\Call{Search}{\var{buckets}, $N$})
{

    $P = \varnothing$ \\
    $\var{positions} = \varnothing$ \\
    allocate the bitmap $\var{taken}[0..N-1]$ with all \bit{0}s \\

    \For{$i = 0; \, i < m; \, i = i + 1$ :}{
        $\var{bucket} = \var{buckets}[i]$ \\
        $k_i = 0$ \Comment{\text{pilot for $i$-th bucket}} \\
        \While{\code{true} :}{

            clear \var{positions} \\
            $j = 0$ \\
            \For{; $j < |\var{bucket}.\var{hashes}|; \, j = j + 1$ :}{
                $p = (\var{bucket}.\var{hashes}[j] \oplus h(k_i)) \mymod{N}$\\
                \If{$\var{taken}[p] = \bit{1}$ :}{
                    $k_i = k_i + 1$ \\
                    \code{break} \Comment{\text{try next pilot}} \\
                }
                add $p$ to \var{positions}\\
            }

            \If{$j = |\var{bucket}.\var{hashes}|$ :}{

                \If{\var{positions} contains duplicates :}{
                    $k_i = k_i + 1$ \\
                    \code{continue} \Comment{\text{try next pilot}} \\
                }

                add $\langle \var{bucket}.\var{id}, k_i \rangle$ to $P$ \Comment{\text{save pilot}} \\
                \code{for} all $p$ in \var{positions} :
                    $\var{taken}[p] = \bit{1}$\\

                \code{break}\\
            }

        }
    }

    $\Call{Sort}{P}$ \\
    \code{return} $P$

}

%% file: algs/P_i.tex
\DontPrintSemicolon
\SetAlgoCaptionSeparator{\footnotesize\normalfont\sffamily{.}}
\SetAlgorithmName{\footnotesize\normalfont\sffamily{Alg.}}{}
\SetAlCapFnt{\footnotesize\normalfont\sffamily}
\SetAlCapNameFnt{\footnotesize\normalfont\sffamily}
\SetKwBlock{Begin}{}{}
\Begin($P[i]$)
{
    $\var{block} = \lfloor i / b \rfloor$ \\
    $\var{offset} = i \mymod{b}$ \\
    $w = W[\var{block} + 1] - W[\var{block}]$ \\
    $position = W[\var{block}] \cdot b + \var{offset} \cdot w$ \\
    $x = B[\var{position}, \var{position} + w]$ \\
    \code{return} $x$
}

%% file: impl.tex

\section{Implementation Details}\label{sec:impl}

In this section, we detail concrete implementations
of the {\pth} construction from Section~\ref{sec:main}.
These implementations are meant to deliver good
practical performance, hence,
exploit multi-threading and external-memory processing.
To this end, we first present in Section~\ref{sec:parallel_search}
a fundamental building block -- a \emph{parallel search} procedure.
In Section~\ref{sec:int_mem} we
assume that the whole input set $S$ can be
processed in internal memory.
In Section~\ref{sec:ext_mem}, instead,
we aim at building the {\pth} data structure for very large sets
that cannot be processed in internal memory,
thus it is necessary to use temporary disk storage.

For all constructions,
the pairs $\langle \var{id},\var{hash}\rangle$,
defined during the map step in Section~\ref{sec:map},
consume an integral number of bytes, $q=q_1+q_2$,
where $q_1$ is the number of bytes for \var{id}
and $q_2$ for \var{hash}.
For example, $q = 12$ if $q_1=4$
(allowing up to $2^{32}$ buckets)
and $q_2=8$ (64-bit hash codes).

\subsection{Parallel Search}\label{sec:parallel_search}

As already noted in Section~\ref{sec:main}, the search
step is the core of the whole {\pth} construction
and the most time-consuming step.
Therefore we would like to exploit all the target machine cores
to search pilots more efficiently.
But parallelizing the algorithm in Alg.~\ref{alg:seq_search}
presents some serious limitations
in that we cannot directly compute pilots in parallel,
say, $K$ pilots for $K$ different buckets,
because the displacement of keys
of the $i$-th bucket depends on the displacement
of \emph{all} keys of the previous buckets.

\begin{figure}[t]
\centering
\includegraphics[scale=0.67]{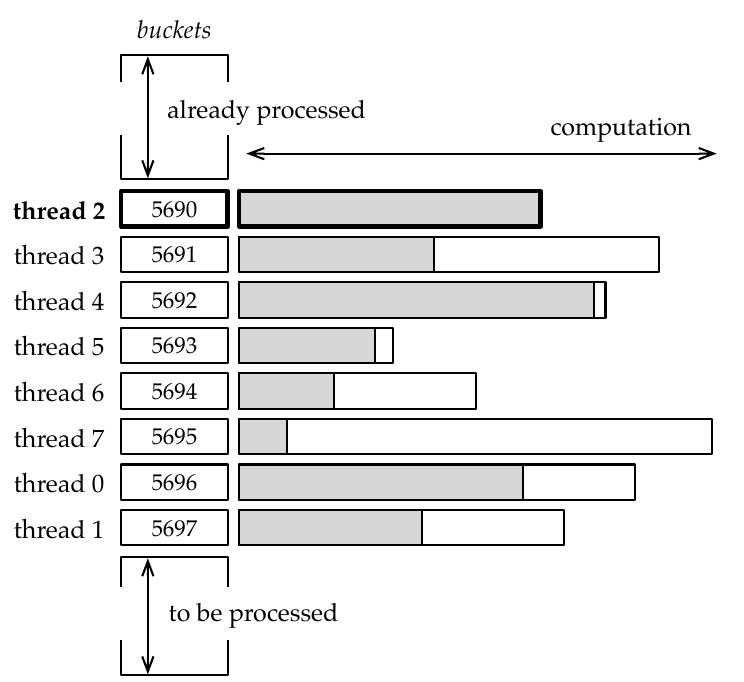}
\mycaption{A graphical representation of the parallel search procedure
with 8 threads, when processing buckets
from index 5690 to index 5697.
The bars to the right of the buckets represent
the amount of computation that threads have to do
before computing the first successful pilot.
The \emph{shaded part} of the bars corresponds to the work done
for \emph{unsuccessful} pilots (those that create collisions in the
bitmap). In this example, thread 2 (in bold font) has to commit its work,
therefore it can execute until success (the shaded part covers its entire bar).
The crucial point is that:
\emph{the shaded part can be computed in parallel}, hence
saving time compared to the sequential implementation.
\label{fig:parallel_search}}
\end{figure}

However, \emph{while a thread may not finalize
the computation of a pilot, it can try
some pilots anyway given the current bit-configuration of the bitmap
and, thus, potentially discard many pilots
that surely cannot work because of the occupied slots.}
Fig.~\ref{fig:parallel_search} illustrates a pictorial
representation of this idea.
Based on this idea, we proceed as follows.

We spawn $K$ parallel threads and
work on $K$ consecutive buckets with
one thread per bucket.
Each thread advances in the search of a pilot
independently,
pausing the search when a pilot that works
for the \emph{current state} of the bitmap is found.
At that point, either the thread is processing
the first bucket to commit, or it is waiting
for some other thread to update the bitmap.
In the former case, the thread (i) concludes the pilot search,
(ii) updates the bitmap,
and (iii) starts processing the next bucket to be processed.
Only one thread at a time can thus update the bitmap.
In the latter case, the waiting thread wakes up after a bitmap update
and continues the pilot search according to the new state of
the bitmap,
pausing as soon as it finds a new working pilot.

The crucial point, now, is to manage the turns
between the threads in an efficient way.
To do so,
we label the threads with unique identifiers, from $0$ to $K-1$,
and maintain the following invariant:
thread $i$ processes the bucket of index $0 \leq j < m$
where $i = j \mymod K$
(see Fig.~\ref{fig:parallel_search} for an example
with $K=8$).
\emph{It follows that the threads must finalize
their respective computation
and commit changes to the bitmap following the
identifier order to guarantee correctness.}
We guarantee this ordering using a \emph{shared} identifier,
periodically indicating the thread that is allowed to commit.
Therefore, a thread advances the search as far as it
can by checking (inside a loop, sometimes called ``busy waiting'')
the shared identifier.
If the shared identifier is equal to its own identifier,
then the thread can commit its work.
Note that we do \emph{not} require a lock/unlock mechanism
to synchronize the threads
that would sensibly erode the benefit of parallelism,
but rely on the \emph{value} assumed by the shared identifier.
This is a rather important point to obtain
good practical performance.

At the beginning of the algorithm, the shared identifier
is equal to $0$, hence only the first thread is allowed
to conclude the search and update the bitmap.
Then, the shared identifier is set to $1$ by the first thread.
The second thread can now commit, and so on.
In general, the turn of the thread $i$ comes when
all threads $0 \leq j < i$ have committed.
The shared identifier is set back to $0$ when the thread $K-1$ commits.

\subsection{Internal Memory}\label{sec:int_mem}

In this setting, we assume that the construction algorithm
has enough internal memory available.
We have to specify only few details
given the flexibility of the algorithm presented
in Section~\ref{sec:main}.

\begin{itemize}

\item The blocks output by the map step are materialized
as in-memory vectors of pairs.
The total space taken is therefore $n(q_1+q_2)$ bytes.
Let $K$ be the number of parallel threads to use.
The map step spawns $K$ threads. Each thread
allocates a vector of $n/K$ pairs and fills it
by forming pairs from a partition of $S$
consisting in $n/K$ keys.
Once the vector is filled up, it is sorted.
Since the threads work on different
partitions of $S$, the map with $K$ threads
achieves nearly ideal speed-up.

\item The collection of buckets output by the merge step
consists in $L$ in-memory vectors of hash codes,
whose total cost is $(n+m)q_2$ bytes.
During the merge procedure, the blocks output by the map
and the buckets under construction co-exist, hence
the total memory usage is $n(q_1+q_2)+(n+m)q_2 = n(q_1+2q_2)+mq_2$ bytes.

There is no meaningful opportunity for
parallelization during the merge step.

\item The auxiliary arrays used by the
search step, i.e., \var{taken} and \var{positions}
are materialized as in-memory arrays of 64-bit integers.
The pilots table $P$ in the pseudocode of Alg.~\ref{alg:seq_search}
is also implemented with a 64-bit integer array.
Since $P$ fits in internal memory, the saving of a pilot
in line 21 simplifies to $P[\var{bucket}.\var{id}] = k_i$
so that $P$ is already sorted and the $\Call{Sort}{P}$ step
in line 24 is \emph{not} performed at all.
The search can be executed in parallel using $K$ threads
as we explained in Section~\ref{sec:parallel_search}.
The total space used for \var{taken} is $\lceil\frac{n}{8\alpha}\rceil$
bytes.
The space for $P$ is $8m$ bytes.
The space for the \var{free} array is $(\lceil\frac{n}{\alpha}\rceil-n)8$
bytes instead.

\end{itemize}

The overall space used by the internal-memory construction
is therefore
\begin{align*}
\max\{ & n(q_1+2q_2)+mq_2 , \\
       & \lceil n/(8\alpha) \rceil + 8m + (\lceil n/\alpha\rceil-n)8 + (n+m)q_2 \}
\end{align*}
bytes. For example, the space is at most
$\max\{ 20n + 8m , 16.25n + 16m \}$
bytes for $\alpha>1/2$, which is $20n+8m$ bytes unless a very large $c$ is used.

\subsection{External Memory}\label{sec:ext_mem}

When not enough internal memory is available
for the construction, e.g., because the
size of $S$ is very large, it is mandatory
to resort to external memory.
The general algorithm described in
Section~\ref{sec:main} easily adapts to this
scenario because
the steps of the algorithm
do \emph{not} change,
rather the output of map and merge is written
to (resp. read from) disk.
Let $M$ indicate the maximum amount of internal memory
(in bytes)
that the construction is allowed to use,
and $q$ be the number of bytes taken by a
$\langle \var{id},\var{hash}\rangle$ pair.

\begin{itemize}

\item 
During the map step,
we allocate an in-memory vector of size $\lceil M/q \rceil$ pairs.
Whenever the vector fills up, it is sorted (possibly
in parallel) and flushed to a new file on disk,
as to guarantee that at most $M$ bytes of internal memory
are used during the process.

\item 
Let $K$ be the number of files created during the map step,
i.e., $K = \lceil (qn)/M \rceil$.
The pairs from these $K$ files are merged together
to create the collection of $L$ buffers
representing the buckets.
Again, the $L$ buffers are formed in internal memory
without violating the limit of $M$ bytes. Whenever
the limit is reached, the content of the buffers is
accumulated to $L$ files on disk.

\item The buckets are read from the $L$ files
and processed by the search procedure,
possibly by its parallel implementation
described in Section~\ref{sec:parallel_search}.

We point out that
the bitmap \var{taken} is \emph{always} kept
in internal memory
(and the array \var{positions} as well),
because its access pattern is \emph{random}
and the search would be slowed down to unacceptable
rates if the bitmap were resident on disk.
Therefore, since the search needs
$N = \lceil n/\alpha \rceil$ bits of internal memory,
we are left with $M - \lceil N/8 \rceil$ bytes
available to store the pilots.

The pilots are accumulated in an in-memory vector
of $\lceil (M - \lceil N/8 \rceil)/q \rceil$ pairs $\langle \var{id}, \var{pilot} \rangle$
(also these pairs consume $q$ bytes each as those
used during the map step).
Whenever the $M$-byte limit is reached,
the vector is flushed to disk
and emptied. Lastly, the different files containing
the pilots are merged together to obtain the final
$P$ table -- the $\Call{Sort}{P}$ step in line 24
of the algorithm in Alg.~\ref{alg:seq_search}.

\end{itemize}

%% file: experiments.tex

\section{Experimental Results}\label{sec:experiments}



For the experimental results in this section,
we use a server machine equipped
with 8 Intel i9-9900K cores (@3.60 GHz),
64 GB of RAM DDR3 (@2.66 GHz), and running Linux 5 (64 bits).
For all experiments where
parallelism is enabled \emph{we use 8 parallel threads},
one thread per core.
Each core has two private levels of cache memory:
32 KiB L1 cache (one for instructions and one for data);
256 KiB for L2 cache. A shared L3 cache spans 16 MiB.
All cache levels have a line size of 64 bytes.
All datasets are read from a
Western Digital Red mechanical disk with
4 TB of storage and a rotation rate of 5400 rpm
(SATA 3.1, 6 GB/s).

The implementation of {\pth}
is written in C++ and available at
\url{https://github.com/jermp/pthash}.
We use the C++ \texttt{std::thread} library to support parallel execution,
i.e., without relying on other frameworks such as Intel's TBBs or OpenMP
that would otherwise limit the portability of our software.
For the experiments reported in the article,
the code was compiled with gcc 9.2.1
using the flags:
\texttt{-std=c++17 -pthread -O3 -march=native}.

\subsection*{Methodology}
Construction time is reported in
total seconds (resp., minutes in external memory), taking the average between 3 runs.
Lookup time is measured by loading the MPHF data structure in memory and looking up every single
key in the input using a single core of the processor.
The reported lookup time is the average between 5 runs.
For inputs residing on disk, we load
batches of 8 GB of keys in internal memory
and measure lookup time on each batch.
The final reported time is the weighted average among
the measurements on all batches.
Lookup time is reported in nanoseconds per key ({\nspk}).
The space of the MPHF data structures is reported
in bits per key ({\bpk}).

A testing detail of particular importance
is that, before running a construction algorithm,
\emph{the disk cache is cleared} to ensure that the whole
input dataset is read from the disk.

\begin{table}[t]
\centering
\caption{String collections used in the experiments.\vspace{-3mm}}
\scalebox{\mytablescale}{\input{tables/string_collections.tex}}
\label{tab:string_collections}
\end{table}


\subsection*{Datasets}
We use some real-world string collections
as input datasets.
Table~\ref{tab:string_collections} reports
the basic statistics for these collections.
All datasets are publicly available for
download by following the corresponding
link in the References.


We use natural-language $q$-grams
as they are in widespread use in IR, NLP, and Machine-Learning
applications.
Specifically, we use the 2-3-grams from the English GoogleBooks
corpus, version 2~\cite{GoogleBooksV2},
as also used in prior work on the problem~\cite{belazzougui2014cache,limasset2017fast}.
URLs are interesting as they represent a sort of ``worst-case'' input
given their very long average length.
We use those of the Web pages
in the ClueWeb09 dataset~\cite{ClueWeb09}
(category B and full dataset),
and those collected in 2005
from the UbiCrawler~\cite{boldi2004ubicrawler}
relative to the .uk domain~\cite{UK2005}.
TweetsKB~\cite{fafalios2018tweetskb}
is a collection of unique tweets identifiers (IDs),
corresponding to tweets collected from February 2013
to December 2020.


As TweetsKB IDs, ClueWeb09-Full URLs, and GoogleBooks 3-gr do not fit in the internal
memory of our test machine (64 GB), we are going to use these
three datasets to benchmark the construction algorithms
in external memory.
The other three datasets can be processed
in internal memory.

\begin{table*}[t]
\centering
\caption{\textbf{Internal-memory} construction time (in seconds),
space, and lookup time, for a range of algorithms.\\
Numbers in parentheses refer to the parallel construction using 8 threads.
All {\pth} configurations use $\alpha=0.94$ and $c=7.0$.
\vspace{-3mm}}
\scalebox{\mytablescale}{\input{tables/int_mem.tex}}
\label{tab:int_mem}
\end{table*}

\subsection*{Algorithms and Implementations}
{\pth} is compared against the
state-of-the-art algorithms reviewed in Section~\ref{sec:related_work}.
Table~\ref{tab:construction_bounds} at page~\pageref{tab:construction_bounds}
indicates what
implementations support parallel execution and
external-memory processing.
The ``HEM'' suffix used in the tables
stands for \emph{heuristic external memory}~\cite{botelho2013practical}
and refers to the approach of partitioning
the input and building an independent MPHF
on each partition.

Whenever possible, we used the implementations available from
the original authors. We include a link to the source code in the
References section.
All implementations are in C/C++, except for the construction of GOV
which is only available in Java
(but lookup time is measured using a C program).
Moreover, we also tested the algorithms \emph{using
the same parameters as suggested by their respective authors}
to offer different trade-offs
between construction time and space effectiveness.
Below we report some details.


FCH
is the only algorithm that we re-implemented (in C++) faithfully
to the original paper.
%
For the CHD
algorithm we were unable to use $\lambda=7$
for more than a few thousand keys.
%
The EMPHF library
also includes the corresponding HEM implementation
of the algorithm, EMPHF-HEM.
%
%
%
The authors of BBHash also considers $\gamma=5.0$
in their own work
but we obtained a space larger than
6.8 {\bpk}, so we excluded this value of $\gamma$
from the analysis.
%



\subsection{Internal Memory}\label{sec:int_mem_exp}

Table~\ref{tab:int_mem} shows the
performance of the algorithms on the datasets that can be
processed entirely in internal memory.
The numbers in parentheses refer to the parallel
construction using 8 threads; moreover,
all {\pth} configurations use $\alpha=0.94$
and $c=7.0$.

Let us first consider the sequential construction
and formulate some general observations.
{\pth} is the fastest MPHF data structure at evaluation time.
FCH offers similar lookup performance but {\pth} is more
space-efficient and much faster at construction time
(FCH with $c=3.0$ takes too much time to run over the GoogleBooks 2-gr
dataset).
Moreover, {\pth} offers good space effectiveness, albeit not the best:
around 3.2 {\bpk} using the D-D encoding but less using PC,
i.e., around 2.7 -- 2.8 {\bpk}, and even less using Elias-Fano (EF),
i.e., around 2.4 -- 2.5 {\bpk}.
For algorithms achieving a similar space, such as BBHash,
{\pth} is 2$\times$ faster at lookup time, and even better at construction time.
Compared to more space-efficient algorithms,
such as RecSplit or CHD, {\pth} is 2 -- 6$\times$ faster at lookup time
with better construction time.

For the HEM implementation of {\pth}
we fix the average partition size to
$b = 5 \times 10^6$ and let the
number of partitions be approximately $n/b$.
Therefore, we use 8, 16, and 128 partitions for
UK2005 URLs, ClueWeb09-B URLs, and GoogleBooks 2-gr
respectively.
It is worth noting that {\pth}-HEM is faster to build than {\pth},
and the gap increases by increasing $n$.
For example, building 128 partitions
takes 170 seconds on the GoogleBooks 2-gr dataset
instead of 410 seconds, hence 2.45$\times$ faster.
This is a direct consequence of the complexity of the search
that is \emph{not} linear in $n$
as per Theorem~\ref{thm:search}.
Thus, building $n/b$ functions each for $\approx$$b$ keys
is faster than building a \emph{single} function for $n$ keys.

Another meaningful point to observe is that
{\pth}-HEM imposes a slight penalty at lookup time
compared to {\pth} (e.g., around 10 -- 16\% on most datasets and encoders),
with no space overhead.
As a comparison, EMPHF-HEM imposes a penalty of 17 -- 36\% at lookup time and of 26 -- 33\% of space usage compared to EMPHF.
It is desirable that
the space of the HEM representation
with $r$ partitions
is very similar to that of the un-partitioned data structure
built from the same input
with $m = \lceil cn / \log_2(n) \rceil$ buckets.
To this end, we search for each MPHF partition
using exactly $\lfloor m/r \rfloor$ buckets
so to guarantee that the total number of buckets
used by the HEM data structure is $m$.
In fact, note that using $\lfloor m/r \rfloor$ buckets for
the $i$-th partition
is different than letting the search over $n_i$ keys
to use $\lceil c n_i / \log_2(n_i) \rceil$ buckets
because of the non-linearity of the $\log_2$ function.
In particular, for an average partition size $n_i$ close to $n/r$,
$\lfloor m/r \rfloor < \lceil c n_i / \log_2(n_i) \rceil$ holds.

We now discuss the results for parallel constructions.
Since a parallel construction has the clear advantage
of reducing construction time, it can be used
to either:
\begin{itemize}
\item build functions more quickly, for a \emph{fixed space budget};
\item build more compact functions
by selecting tighter parameters $c$ and $\alpha$ for the search,
for a \emph{fixed construction-time budget}.
\end{itemize}

\noindent
The numbers in parentheses in Table~\ref{tab:int_mem}
show that construction time improves significantly.
The parallel algorithm reduces construction time
by 2 -- 2.5$\times$, thus allowing {\pth}
to build faster than its direct competitors
that also exploit multi-threading, like GOV and BBHash.
Better speedup can be achieved, not surprisingly,
by {\pth}-HEM because partitions can be built
independently in parallel.

\begin{table*}[t]
\centering
\caption{\textbf{External-memory} construction time (in {minutes}),
space, and lookup time, for a range of algorithms.\\
Numbers in parentheses refer to the parallel construction using 8 threads.
All {\pth} configurations use $\alpha=0.94$ and $c=7.0$.
\vspace{-3mm}}
\scalebox{\mytablescale}{\input{tables/ext_mem.tex}}
\label{tab:ext_mem}
\end{table*}

\begin{figure}[t]
\centering
\includegraphics[scale=0.5]{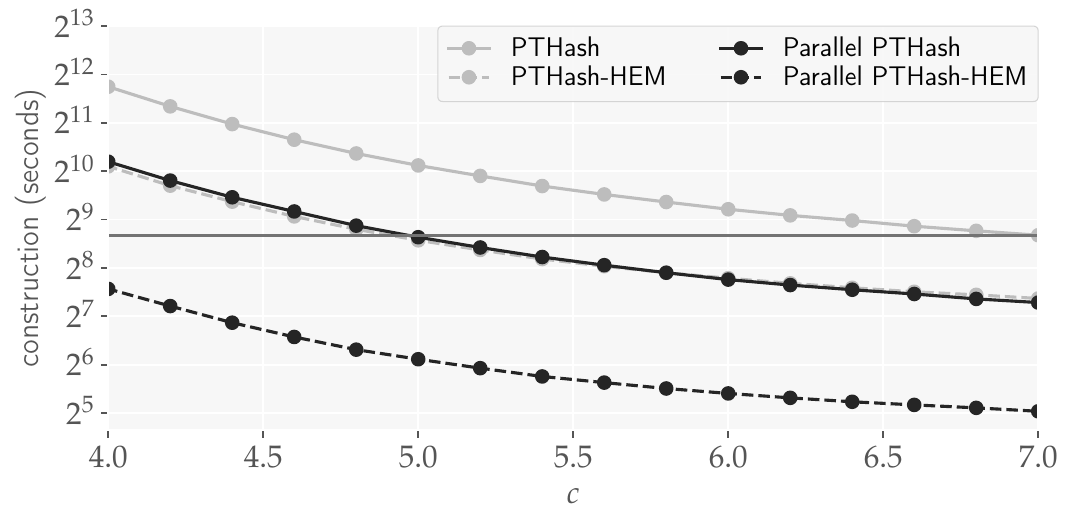}
\vspace{-7mm}
\mycaption{Internal-memory construction time of {\pth} and {\pth}-HEM on the GoogleBooks 2-gr dataset, by varying $c$,
and for $\alpha=0.94$.
The parallel construction uses 8 threads.
The horizontal line marks the lowest construction time achieved
by sequential {\pth}, that is for $c=7.0$.
Note the log-scale of the y-axis.
\label{fig:constr}}
\end{figure}

\begin{figure}[t]
\centering
\includegraphics[scale=0.5]{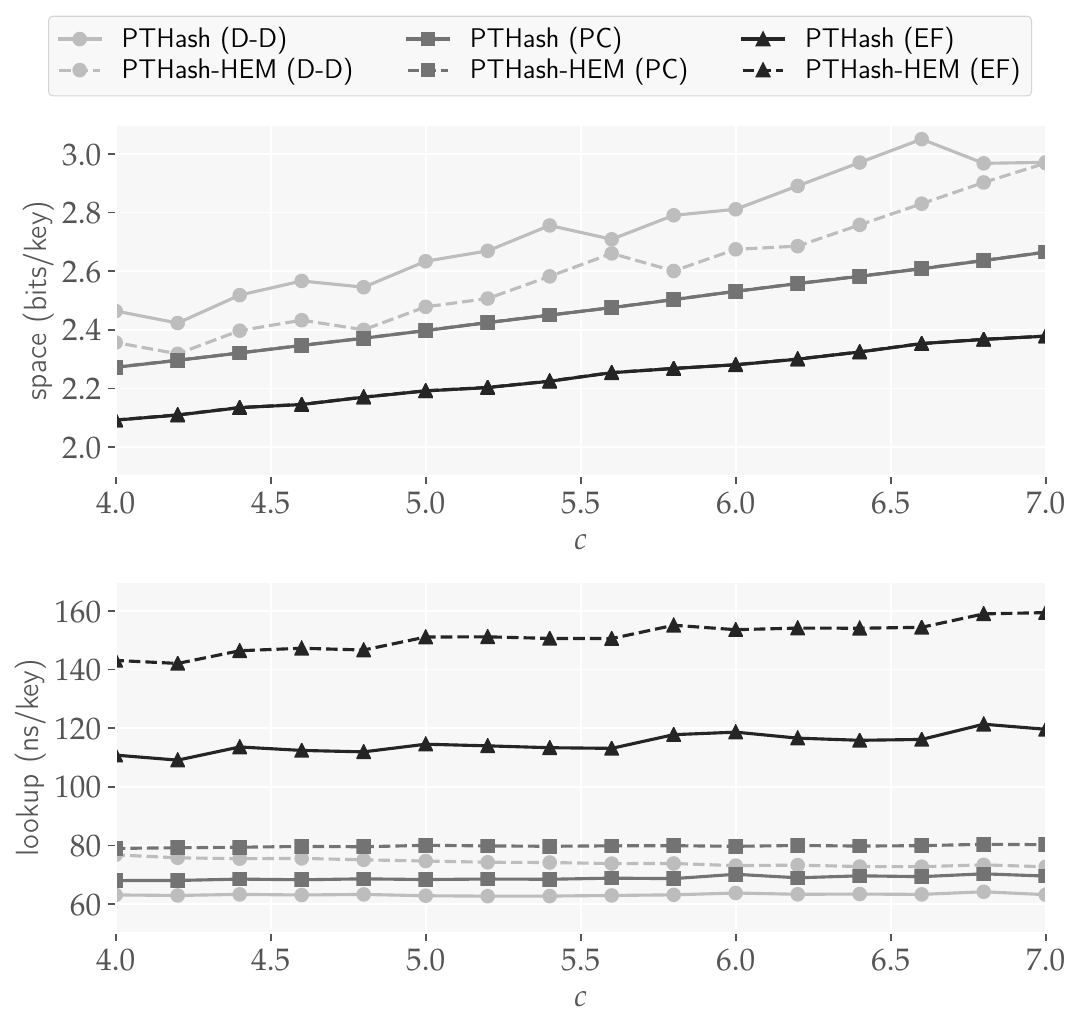}
\vspace{-7mm}
\mycaption{Space/time trade-offs for {\pth} and {\pth}-HEM by varying $c$
and compressor (D-D, PC, EF), and for $\alpha=0.94$.
Performance was measured on the GoogleBooks 2-gr dataset.
\label{fig:space_time_tradeoff}}
\end{figure}

Fig.~\ref{fig:constr} highlights, instead,
that parallelism can also be used
to build more space-efficient functions.
In the figure, the horizontal line marks the
lowest construction time achieved by sequential {\pth}
(for $c=7.0$), so that it is easy to see the \emph{same}
time is spent by the parallel {\pth} for a \emph{smaller} $c$.
For example, $c$ can be reduced from 7.0 to 5.0
for $\alpha=0.94$ 
to build a smaller MPHF within the same time-budget using 8 parallel threads.
The companion Fig.~\ref{fig:space_time_tradeoff}
shows the space consumption and lookup time
corresponding to such values of $c$.
Continuing the same example, for the \mbox{D-D} encoder,
space improves
from 2.97 to 2.63 {\bpk} for $\alpha=0.94$.
As another example (not shown in the plots),
for $\alpha=0.99$ we obtained a similar space improvement:
from 2.89 to 2.52 {\bpk}.

Fig.~\ref{fig:space_time_tradeoff} also shows that
the partitioned compact encoding (PC)
introduced in Section~\ref{sec:main}
is always more effective than \mbox{D-D} with only
a minor lookup penalty.
PC may be preferable to EF for its better
lookup performance despite
the latter being still more space-efficient.
We report two more illustrative examples:

\begin{itemize}

\item Compared to RecSplit $(\ell=12,b=9)$,
parallel {\pth} with $\alpha=0.99$ and $c=4.0$
builds faster (2710 vs. 3789 seconds)
and, using PC, it retains even better space (2.12 vs. 2.23 {\bpk}),
with $3.75\times$ faster lookup time.

\item Elias-Fano (EF) allows {\pth} to break the 2 {\bpk} ``barrier'',
indeed consuming only 1.98 {\bpk} for $\alpha=0.99$ and $c=4.0$:
we point out that, up to date, only RecSplit was able to do so.
Using that configuration of parameters, {\pth}-HEM
is only slightly larger than RecSplit $(\ell=8,b=100)$,
i.e., 1.98 vs. 1.80 {\bpk},
but builds nearly $2\times$ faster
with $1.5\times$ better lookup time.

\end{itemize}

\subsection{External Memory}\label{sec:ext_mem_exp}

Table~\ref{tab:ext_mem}
shows the performance of the algorithm supporting external-memory processing,
on datasets comprising several billions of strings.
For {\pth} and {\pth}-HEM we fixed the amount
of internal memory to 16 GB (out of the 64 GB available)
as we did not
observe any appreciable difference by using less or more memory.
The number of partitions used by {\pth}-HEM
is set using the same methodology of Section~\ref{sec:int_mem_exp},
i.e., we fix the average partition size to $b = 5 \times 10^6$
and let the
number of partitions be approximately $n/b$.
Therefore, we use 400, 1000, and 1500 partitions for
TweetsKB IDs, ClueWeb09-Full URLs, and GoogleBooks \mbox{3-gr} respectively.

Results for external memory are \emph{consistent}
with those discussed for internal memory:
the construction time of {\pth} is competitive
or better than that of other algorithms,
for similar space effectiveness
yet better lookup time.

We point out, again, that the parallel search algorithm
described in Section~\ref{sec:parallel_search}
is a key ingredient to achieve efficient parallel
construction, even in external memory.
{\pth}-HEM offers further reductions in construction
time at the cost of a lookup penalty.
However, note that the parallel speedup
is partially eroded in external memory
compared to internal memory, due to I/O operations,
even though these are just \emph{sequential} reads/writes.

Considering Table~\ref{tab:ext_mem}, we now
illustrate some comparisons using the largest
dataset as example, i.e., GoogleBooks 3-gr with approximately 7.4 billion strings,
noting that very similar considerations are valid
for the other datasets:

\begin{itemize}

\item {\pth} with D-D is $2\times$ faster than BBHash with $\gamma=1.0$
at construction (163 vs. 337 seconds),
retains even better space usage (2.91 vs. 3.07 {\bpk}),
and is $3\times$ faster at lookup time (91 vs. 242 {\nspk}).

\item {\pth} with D-D is slightly faster than GOV at construction time
with $2.7\times$ faster lookup (91 vs. 242 {\nspk}),
albeit being less space-efficient (2.91 vs. 2.23 {\bpk}).
Instead, {\pth}-HEM with EF is as efficient as GOV
regarding space and lookup time, but with almost $2\times$
better construction time (87 vs. 180 seconds).

\item {\pth} with PC is nearly $4\times$ faster at construction
than EMPHF (163 vs. 629 seconds),
retains the same space and $1.5\times$ better lookup (143 vs. 220 {\nspk}).
The HEM implementation of {\pth} outperforms the HEM implementation
of EMPHF under every aspect.

\end{itemize}

%% file: tables/string_collections.tex
\begin{tabular}{lrrc}

\toprule

Collection & num. strings & GBs & avg. str. length \\
\midrule


UK2005 URLs          &    39,459,925  &   2.9  &   71.37 \\
ClueWeb09-B URLs     &    49,937,704  &   2.8  &   54.72 \\
GoogleBooks 2-gr     &   665,752,080  &  12.1  &   17.21 \\
TweetsKB IDs         & 1,983,291,944  &  39.2  &   18.75 \\
ClueWeb09-Full URLs  & 4,780,950,911  & 326.5  &   67.28 \\
GoogleBooks 3-gr     & 7,384,478,110  & 179.5  &   23.30 \\

\bottomrule
\end{tabular}


%% file: tables/int_mem.tex
\setlength{\tabcolsep}{2pt}
\begin{tabular}{
l c rlcc c rlcc c rlcc}

\toprule

\multirow{3}{*}{Algorithm}

&& \multicolumn{4}{c}{UK2005 URLs}
&& \multicolumn{4}{c}{ClueWeb09-B URLs}
&& \multicolumn{4}{c}{GoogleBooks 2-gr} \\

\cmidrule(lr){3-6}
\cmidrule(lr){7-12}
\cmidrule(lr){13-16}

&& \multicolumn{2}{c}{construction} & space & lookup
&& \multicolumn{2}{c}{construction} & space & lookup
&& \multicolumn{2}{c}{construction} & space & lookup \\

&& \multicolumn{2}{c}{(\secs)} & ({\bpk}) & ({\nspk})
&& \multicolumn{2}{c}{(\secs)} & ({\bpk}) & ({\nspk})
&& \multicolumn{2}{c}{(\secs)} & ({\bpk}) & ({\nspk}) \\

\midrule

{\pth} (D-D)
&& \multirow{3}{*}{9} & \multirow{3}{*}{(5)} & 2.82 & 49
&& \multirow{3}{*}{12} & \multirow{3}{*}{(6)} & 3.07 & 49
&& \multirow{3}{*}{410} & \multirow{3}{*}{(156)} & 2.97 & 63 \\

{\pth} (PC)
&&&& 2.82 & 49
&&&& 2.80 & 52
&&&& 2.67 & 69 \\

{\pth} (EF)
&&&& 2.50 & 59
&&&& 2.49 & 63
&&&& 2.38 & 121 \\

\midrule

{\pth-HEM} (D-D)
&& \multirow{3}{*}{8} & \multirow{3}{*}{(2)} & 3.11 & 52
&& \multirow{3}{*}{10} & \multirow{3}{*}{(2)} & 3.08 & 57
&& \multirow{3}{*}{167} & \multirow{3}{*}{(34)} & 2.97 & 72 \\

{\pth-HEM} (PC)
&&&& 2.82 & 54
&&&& 2.80 & 59
&&&& 2.67 & 80 \\

{\pth-HEM} (EF)
&&&& 2.50 & 66
&&&& 2.49 & 70
&&&& 2.38 & 159 \\

\midrule

FCH ($c=3.0$)
&& 1138 & & 3.00 & 52
&& 1438 & & 3.00 & 55
&& -- & & -- & -- \\

FCH ($c=4.0$)
&& 266 & & 4.00 & 60
&& 267 & & 4.00 & 64
&& 9110 & & 4.00 & 54 \\

FCH ($c=5.0$)
&& 119 & & 5.00 & 68
&& 107 & & 5.00 & 71
&& 3225 & & 5.00 & 55 \\

\midrule

CHD ($\lambda=4$)
&& 32 & & 2.17 & 170
&& 43 & & 2.17 & 185
&& 1251 & & 2.17 & 410 \\

CHD ($\lambda=5$)
&& 84 & & 2.07 & 173
&& 115 & & 2.07 & 182
&& 3923 & & 2.07 & 410 \\

CHD ($\lambda=6$)
&& 306 & & 2.01 & 173
&& 429 & & 2.01 & 178
&& 15583 & & 2.01 & 406 \\

\midrule

EMPHF
&& 10 & & 2.61 & 126
&& 13 & & 2.61 & 135
&& 276 & & 2.61 & 211 \\


EMPHF-HEM
&& 9 & & 3.49 & 152
&& 11 & & 3.30 & 158
&& 148 & & 3.44 & 287 \\

\midrule

GOV
&& 39 & (14) & 2.23 & 87
&& 47 & (15) & 2.23 & 94
&& 613 & (170) & 2.23 & 170 \\

\midrule

BBHash ($\gamma=1.0$)
&& 50 & (12) & 3.10 & 154
&& 66 & (14) & 3.06 & 174
&& 1248 & (189) & 3.08 & 273 \\

BBHash ($\gamma=2.0$)
&& 31 & (7) & 3.71 & 133
&& 40 & (9) & 3.71 & 150
&& 666 & (102) & 3.71 & 204 \\

\midrule

RecSplit ($\ell=5$, $b=5$)
&& 6 & & 2.95 & 153
&& 8 & & 2.95 & 148
&& 132 & & 2.95 & 244 \\

RecSplit ($\ell=8$, $b=100$)
&& 37 & & 1.79 & 113
&& 47 & & 1.79 & 118
&& 724 & & 1.80 & 202 \\

RecSplit ($\ell=12$, $b=9$)
&& {225} & & {2.23} & {93}
&& {284} & & {2.23} & {107}
&& {3789} & & {2.23} & {225} \\

\bottomrule
\end{tabular}

%% file: tables/ext_mem.tex
\setlength{\tabcolsep}{2pt}
\begin{tabular}{
l c rlcc c rlcc c rlcc}

\toprule

\multirow{3}{*}{Algorithm}

&& \multicolumn{4}{c}{TweetsKB IDs}
&& \multicolumn{4}{c}{ClueWeb09-Full URLs}
&& \multicolumn{4}{c}{GoogleBooks 3-gr} \\

\cmidrule(lr){3-6}
\cmidrule(lr){7-12}
\cmidrule(lr){13-16}

&& \multicolumn{2}{c}{construction} & space & lookup
&& \multicolumn{2}{c}{construction} & space & lookup
&& \multicolumn{2}{c}{construction} & space & lookup \\

&& \multicolumn{2}{c}{(minutes)} & ({\bpk}) & ({\nspk})
&& \multicolumn{2}{c}{(minutes)} & ({\bpk}) & ({\nspk})
&& \multicolumn{2}{c}{(minutes)} & ({\bpk}) & ({\nspk}) \\

\midrule

{\pth} (D-D)
&& \multirow{3}{*}{36} & \multirow{3}{*}{(18)} & 3.07 & 80
&& \multirow{3}{*}{121} & \multirow{3}{*}{(82)} & 2.96 & 120
&& \multirow{3}{*}{163} & \multirow{3}{*}{(98)} & 2.91 & 91 \\

{\pth} (PC)
&&&& 2.61 & 101
&&&& 2.58 & 175
&&&& 2.56 & 143 \\

{\pth} (EF)
&&&& 2.36 & 188
&&&& 2.32 & 214
&&&& 2.31 & 208 \\

\midrule

{\pth-HEM} (D-D)
&& \multirow{3}{*}{17} & \multirow{3}{*}{(10)} & 2.85 & 116
&& \multirow{3}{*}{78} & \multirow{3}{*}{(61)} & 2.75 & 152
&& \multirow{3}{*}{87} & \multirow{3}{*}{(59)} & 2.71 & 135 \\

{\pth-HEM} (PC)
&&&& 2.61 & 128
&&&& 2.58 & 192
&&&& 2.57 & 190 \\

{\pth-HEM} (EF)
&&&& 2.36 & 201
&&&& 2.32 & 235
&&&& 2.31 & 230 \\

\midrule

EMPHF
&& 68 & & 2.61 & 207
&& 415 & & 2.61 & 231
&& 629 & & 2.61 & 220 \\

EMPHF-HEM
&& 16 & & 3.03 & 279
&& 67 & & 3.31 & 304
&& 94 & & 3.06 & 304 \\

\midrule

GOV
&& 45 & (25) & 2.23 & 192
&& 138 & (90) & 2.23 & 232
&& 180 & (108) & 2.23 & 242 \\

\midrule

BBHash ($\gamma=1.0$)
&& 80 & (16) & 3.07 & 294
&& 323 & (307) & 3.07 & 320
&& 337 & (160) & 3.07 & 305 \\

BBHash ($\gamma=2.0$)
&& 40 & (10) & 3.71 & 217
&& 185 & (173) & 3.71 & 236
&& 171 & (91) & 3.71 & 235 \\

\bottomrule
\end{tabular}

%% file: conclusions.tex
\section{Conclusions}\label{sec:conclusions}

In this work we described a construction
algorithm for {\pth} that enables
multi-threading and external-memory processing.
These two features are very important
to scale to large datasets in reasonable time.
Our C++ implementation is publicly available at \url{https://github.com/jermp/pthash}.

We presented an extensive experimental analysis
and comparisons,
using large real-world string collections.
There are three efficiency aspects for
minimum perfect hash data structures:
construction time,
space consumption, and lookup time.
\emph{PTHash can be tuned to match the
performance of another algorithm
on one of the three aspects and, then,
outperform the same algorithm on the
other two aspects.}
We remark that lookup time
may be the most relevant aspect
in concrete applications of minimal perfect hashing.
In this regard, {\pth} offers very fast evaluation
being from $2$ to $6\times$ faster than other competitive techniques.

%% file: proof_expected_runtime.tex

In this appendix we prove Theorem~\ref{thm:search}.
To do so, we first introduce some technical machinery.

The thresholds $p_1=0.6n$ and $p_2=0.3m$ introduced in Section~\ref{sec:map}
logically partition the $n$ keys of $S$ into two sets of buckets, $M_1$ and $M_2$:
$M_1$ contains $n_1=0.6n$ keys and $m_1=0.3m$ buckets;
$M_2$ instead contains $n_2=0.4n$ keys and $m_2=0.7m$ buckets.
The number of keys of the different buckets can be modeled by $m$ random variables.
Clearly, the random variables of buckets in $M_1$
are \emph{independent} from those of the buckets in $M_2$.
Moreover, all random variables belonging to the same set of buckets
are distributed in the same way: we
indicate with $X_1$ the bucket size of the buckets from $M_1$
and with $X_2$ that of the buckets from $M_2$.
For $k \in \{1,2\}$, we know that $X_k$ is a binomially-distributed random variable -- $X_k\sim\Binomial(n_k,1/m_k)$ --
with expectation $\lambda_k=n_k/m_k$.
Let $\lambda=n/m=\log_2(n)/c$.
Then $\lambda_1=2\lambda$ and $\lambda_2=(4/7)\lambda$.
We also know that $\Binomial(N,p)$
converges to $\Poisson(\lambda=Np)$ for $N,1/p \to \infty$.
Hence, for $n \to \infty$, we have
$X_k\sim\Poisson(\lambda_k)$
with
\begin{equation}
\label{eq:poisson_approximation}
\Probability(X_k=t)=\frac{e^{-\lambda_k}\lambda_k^t}{t!}
\end{equation}
for $t \in \mathbb{N} \cup \{0\}$ and $k \in \{1,2\}$.


\begin{lemma}\label{lem:loadfactor}
Let $\alpha_{\geq t}$ be the load factor after all buckets of size $\geq t$ have been processed.
Then, for $n \to \infty$, we have
$$
\ExpectedValue[\alpha_{\geq t}] = \alpha \Big( 0.6\Probability(X_1 \geq t-1) + 0.4\Probability(X_2 \geq t-1) \Big).
$$
\end{lemma}
\begin{proof}
Let $n_{\geq t}$ be the number of keys belonging to buckets of size $\geq t$.
Clearly, $\alpha_{\geq t} = \alpha \cdot n_{\geq t}/n$,
thus, by linearity of expectation,
it suffices to compute $\ExpectedValue[n_{\geq t}]$ to prove the lemma.
We can write $n_{\geq t}$
as
$$
n_{\geq t} = \sum_{i=0}^{m_1-1} \sum_{j=t}^{n} j I_{1,ij} + \sum_{i=0}^{m_2-1} \sum_{j=t}^{n} j I_{2,ij}
$$
where
$$
I_{k,ij} =
   \left\{
\begin{array}{ll}
      1 & \text{bucket } i \text{ from partition } k \text{ contains } j \text{ keys} \\
      0 & \text{otherwise} \\
\end{array}
\right.
$$
for $k \in \{1,2\}$.
Now, by linearity of expectation and knowing that $\ExpectedValue[I_{k,ij}]=\Probability(X_k=j)$ for any $i$,
we have that
$$
\ExpectedValue[n_{\geq t}] = m_1 \sum_{j=t}^{n} j\Probability(X_1=j) + m_2 \sum_{j=t}^{n} j\Probability(X_2=j).
$$
For Equation~\ref{eq:poisson_approximation}, we have that
\begin{align*}
& \sum_{j=t}^{n} j \Probability(X_k=j) =
  \sum_{j=t}^{n} j \cdot \frac{e^{-\lambda_k}\lambda^{j}_k}{j!} = \\
& \lambda_k \sum_{j=t}^{n} \frac{e^{-\lambda_k}\lambda^{j-1}_k}{(j-1)!} =
  \lambda_k \sum_{j=t-1}^{n-1} \Probability(X_k=j) = \\
& \lambda_k \Probability(X_k \geq t-1)
\end{align*}
for $k \in \{1, 2\}$.
Replacing the latter equality in the
previous expression for $\ExpectedValue[n_{\geq t}]$,
it follows that
\begin{align*}
\ExpectedValue[n_{\geq t}] & = m_1 \lambda_1 \Probability(X_1 \geq t-1) + m_2 \lambda_2 \Probability(X_2 \geq t-1) \\
& = 0.6 n \Probability(X_1 \geq t-1) + 0.4 n \Probability(X_2 \geq t-1).
\end{align*}
\end{proof}

\begin{lemma}\label{lem:concentration}
The random variable $\alpha_{\geq t}$ is tightly concentrated around $\ExpectedValue[\alpha_{\geq t}]$
with high probability.
\end{lemma}
\begin{proof}
We show the lemma using the McDiarmid's inequality (which, in turn, is a variation of the
Azuma-Hoeffding inequality)~\cite[Theorem 13.7]{10.5555/3134214}.

The variable $\alpha_{\geq t}$ is a random variable that depends on the random
placement of keys into buckets. Let $Y_i$ be the random variable indicating
the bucket into which the \mbox{$i$-th} key falls.
Clearly, $Y_1,\ldots,Y_n$ are independent random variables.
We claim that $\alpha_{\geq t} = f(Y_1,\ldots,Y_n)$ satisfies the Lipschitz condition
with bound $\frac{\alpha t}{n}$. In fact, the placement of a key in a bucket
cannot change the load factor due to buckets of size $\geq t$ by more than $\frac{\alpha t}{n}$
because:
(i) if the key is placed in a bucket of size $\leq t-2$, the load factor does not change;
(ii) if it falls in a bucket of size $t-1$, the load factor increases by $\frac{\alpha t}{n}$;
(iii) lastly, if it falls in a bucket of size $\geq t$,
the load factor increases by $\frac{\alpha}{n}$.
We can therefore apply the McDiarmid's inequality to show that
$$
\Probability(|\alpha_{\geq t}-\ExpectedValue[\alpha_{\geq t}]| \geq \varepsilon)
  \leq 2e^{-2n(\frac{\varepsilon}{\alpha t})^2}, \text{ for } t=1,\ldots,L.
$$
The bound above is at most $2e^{-O(n/(\log_2(n))^2)}=O(1/n)$ since
$t \leq L = \Theta(\log(n)/c)$.
\end{proof}


\begin{lemma}[Theorem 5.4~\cite{10.5555/3134214}]\label{lem:tools}
Let $X\sim\Poisson(\lambda)$.
We have that
$\Probability(X \geq t) \leq e^{-\lambda} (\frac{e\lambda}{t})^t, \text{ for any } t \geq \lambda$.
\end{lemma}

We are now ready to prove Theorem~\ref{thm:search}.

\begin{proof}[Proof of Theorem 1]
As shown by Equation~\ref{fact:num_trials}, the expected number of trials for the $i$-th bucket is
\begin{equation}
\setcounter{equation}{1}
\ExpectedValue[k_i]+1 = 1/p_i = (1 - \alpha(i))^{-|B_i|}.
\end{equation}
Note that, for $|B_i|=t$,
$
(1 - \alpha(i))^{-|B_i|} \leq (1 - \alpha_{\geq t})^{-t}.
$
Here, we can use $\ExpectedValue[\alpha_{\geq t}]$ in place of
$\alpha_{\geq t}$ since we know that the
error committed will be arbitrarily small with high probability from Lemma~\ref{lem:concentration}.

We now separately analyze the large and the small buckets to show an upper bound to the expected number of trials
for each bucket.
When $t \geq 4\lambda+1$ (large buckets), the expected number of trials is at most
$
  (1-\alpha_{\geq t})^{-t}
  \leq (1-\alpha(0.6 \cdot e^{-2\lambda}(\frac{e2\lambda}{t-1})^{t-1} + 0.4 \cdot e^{-(4/7)\lambda}(\frac{e4/7\lambda}{t-1})^{t-1}))^{-t}
$
for Lemma~\ref{lem:tools}.
Note that the latter bound is decreasing in $t$, hence it is at most
$
    (1-\alpha(0.6 \cdot e^{-2\lambda}(\frac{e 2 \lambda}{4 \lambda})^{4 \lambda} + 0.4 \cdot e^{-(4/7)\lambda}(\frac{e 4/7 \lambda}{4 \lambda})^{4 \lambda}))^{-(4 \lambda + 1)} < 6.
$
In this case, the expected number of trials is therefore bounded by a constant.
Instead, when $t \leq 4\lambda$ (small buckets), Equation~\ref{fact:num_trials} can be upper bounded by $(1-\alpha)^{-4\lambda}$.
Therefore, the expected number of trials is at most
$n^{-4/c \cdot \log_2(1-\alpha)} = n^{\Theta(\alpha/c)}$, for $0 < \alpha < 1$.

Taking into account that each trial takes $O(|B_i|)$ time,
the total expected time of \Call{Search}{} (Alg.~\ref{alg:seq_search}) is
$
\sum_{i=0}^{m-1} O(|B_i|) \cdot (\ExpectedValue[k_i]+1) <
O(n) \cdot n^{\Theta(\alpha/c)} = O(n^{1+\Theta(\alpha/c)}).
$
\end{proof}